\documentclass[aps,prb,twocolumn,showpacs]{revtex4}
\usepackage{graphicx,color}
\usepackage{amsmath}
\usepackage{epsfig}
\usepackage{bm}
\newcommand{\be}{\begin{equation}}
\newcommand{\ee}{\end{equation}}

\newcommand{\beq}{\begin{eqnarray}}
\newcommand{\eeq}{\end{eqnarray}}

\def\H1{\widehat{H}_1}

\begin{document}
\topmargin -1cm
\textheight 24cm
\title{Quantum XX chain with interface}
\author{D.\ Baeriswyl and G.\ Ferraz}
\affiliation{Dept.\ of Physics, University of
Fribourg, CH-1700 Fribourg, Switzerland\\ and International Institute of Physics, UFRN, Natal, Brazil}

\date{\today}

\begin{abstract}

The quantum XX chain - or rather ring - is studied as a toy model of an interface. Two transverse field patterns are used to define the interface, on the one hand a staggered field, on the other hand a step-like configuration, from $-h$ to $+h$. The interface leads to Friedel oscillations and proximity effects, in particular close to the quantum phase transition of the bulk, which is a metal-insulator transition in the fermionic language. The most prominent interface effects appear for odd-numbered rings, for which - in contrast to even chains - the ground state is doubly degenerate. In the regime where the bulk energy spectrum is gapped a level appears close to midgap, with a wave function localized in the region of the interface. The two members of the ground state doublet have two different particle number parities and spin components $S_z=\pm\frac{1}{2}$. They also have different energy levels and thus the degeneracy does not originate from different occupancies of a rigid band structure, but rather from a global symmetry. The unitary transformation linking the two degenerate ground states resembles a Majorana operator. Coherent superpositions of the two states may be suitable candidates for well protected qubits.  

\end{abstract}
\pacs{75.10.Pq}

\maketitle
\section{Introduction}

Spin chains are favorite models for studying fundamental quantum phenomena. Thus the XY chain, where only the $x$ and $y$ components of the spin operators are coupled, has been used for discussing quantum phase transitions \cite{Sachdev, Carollo}, quantum dynamics \cite{Barouch, Mukherjee, Tomka} and quantum entanglement \cite{Vidal, Korepin}. It has also served as an illustrative case displaying the effects of randomness on critical properties \cite{Dershko} and as a handy example for scrutinizing variational ground states \cite{Baeriswyl}.

In this paper we use the model for discussing characteristic effects produced by an interface. A multitude of interesting phenomena can occur at the interface separating two different materials,
such as proximity effects -- the spillover of the order present in one material into the other within a certain spatial region -- Friedel oscillations -- a periodic variation of electronic charge density in a metal close to the interface with an insulator -- or the generation of a two-dimensional electron gas at the interface between a doped and an undoped semiconductor, a prerequisite for the Quantum Hall Effect. Recently metallic -- and even superconducting -- states generated at the interface between a band and a Mott insulator have stimulated a whole new field of research \cite{Mannhart}. Similarly, in topological insulators conducting states are produced at an interface (or at the surface), with rather exotic properties \cite{Hasan}. 

\begin{figure}[ht]
	\centering
	\includegraphics[width=5.5cm]{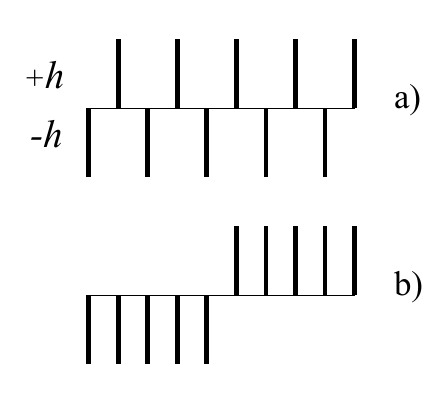}
	\caption[Local magnetization and the susceptibility with the homogeneous field]{\footnotesize (Color online) Examples of inhomogeneous field configurations: a) staggered field, b) step model with a sharp interface.}
	\label{fig:fields}
\end{figure}

We model the interface by choosing particular configurations of the transverse fields, as illustrated in Fig. \ref{fig:fields}. For the staggered field an interface is created by joining the ends of an odd-numbered chain. In the step model the interface is produced by applying opposite fields in the two halves of the chain. 

We consider the isotropic version of the model, the XX chain -- or rather a closed ring -- coupled to a transverse field. The Hamiltonian is
\begin{equation}
H=\sum_{j=1}^L \left\{ J(S_j^xS_{j+1}^x+S_j^yS_{j+1}^y)-h_jS_j^z\right\}\, ,
\label{hamiltonian_1}
\end{equation}
where $S_j^\alpha,\, \alpha=x,y,z$, are spin $\frac{1}{2}$ operators,
$S_{L+1}^\alpha=S_1^\alpha$ and $J$ is chosen positive. The Jordan-Wigner transformation
\beq
S_j^+&=&S_j^x+iS_j^y=c_j^\dag\exp\left(i\pi\sum_{\ell=1}^{j-1}c_\ell^\dag c_\ell\right)\, ,\nonumber \\
S_j^-&=&S_j^x+iS_j^y=\exp\left(-i\pi\sum_{\ell=1}^{j-1}c_\ell^\dag c_\ell\right)c_j
\, ,\nonumber \\
S_j^z&=&c_j^\dag c_j-\frac{1}{2}
\eeq
to spinless fermion operators $c_j, c_j^\dag$ leads to a Hamiltonian of non-interacting particles
\beq
H&=&\frac{J}{2}\sum_{j=1}^{L-1}(c_j^\dag c_{j+1}+c_{j+1}^\dag c_j)
-\sum_{j=1}^L h_jc_j^\dag c_j\nonumber \\
&-&\frac{J}{2}e^{i\pi N}\left(c_L^\dag c_1 +c_1^\dag c_L\right)\, ,
\label{hamiltonian_2}
\eeq
where the particle number $N=\sum_jc_j^\dag c_j$ is related to the $z$ component of the total spin as
\be
S_z=\sum_{j=1}^L \left(c_j^\dag c_j-\frac{1}{2}\right)=N-\frac{L}{2}\, .
\label{spin}
\ee
For odd $N$, the Hamiltonian (\ref{hamiltonian_2}) is a simple tight-binding model with an inhomogeneous chemical potential. For even $N$, the hopping term between sites 
$L$ and $1$ has a different sign than the others. For very large chains this should have negligible effects on bulk properties. However, as we will see later, this boundary term is essential for the ground state degeneracy in odd-numbered chains.

Before discussing interface effects we recall the essential ground-state properties in the absence of an interface. Specifically, we consider the two cases of constant and alternating fields (for an even number of sites in the latter case). Some mathematical details are presented in Appendix
\ref{homogeneous_staggered}.

\begin{figure}[ht]
	\centering
	\includegraphics[width=8cm]{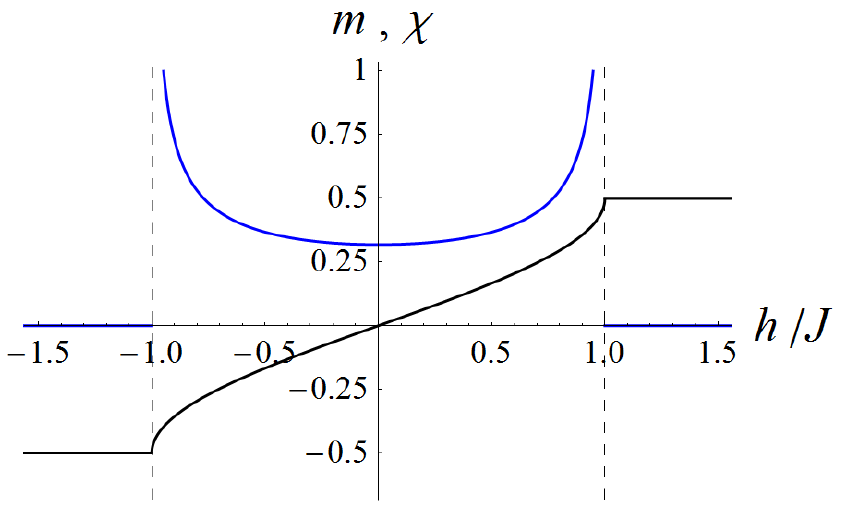}
	\caption{\footnotesize (Color online) Magnetization (lower line) and 
susceptibility (upper line) for a homogeneous transverse field.}
	\label{fig:homogeneous}
\end{figure}

For a constant field, $h_j=h,\quad j=1,...,L$, the main properties of the model have been widely explored, ever since the seminal work of Lieb, Schultz and Mattis \cite{Lieb}. The single-particle spectrum $J\cos k-h$ is negative for all values of the wave vectors, $-\pi\le k\le\pi$, if $h$ exceeds $J$; for $h<-J$, the spectrum is positive. 
For $h=0$ the ground state corresponds to a half-filled band with Fermi wave vectors $\pm \pi/2$. The filling changes as a function of $h$, with Fermi wave vectors moving inwards for $h>0$ and outwards for $h<0$; they disappear at $h=\pm J$, where a Lifshitz transition \cite{Lifshitz} from a partially filled band to either a completely filled or a completely empty band occurs. 
The magnetization per site
\be
m=\langle S_j^z\rangle = \left\{\begin{array}{ll}\frac{1}{2}-\frac{1}{\pi}\arccos{\frac{h}{J}},&0<h<J,\\ 
\frac{1}{2},&h>J\end{array}\right.
\ee
(and correspondingly for $h<0$) is shown in
Fig. \ref{fig:homogeneous}, 
together with the susceptibility $\chi=dm/dh$. The two critical points $h=\pm J$ mark the transitions from a partial magnetization to fully aligned moments. 

\begin{figure}[ht]
	\centering
	\includegraphics[width=8cm]{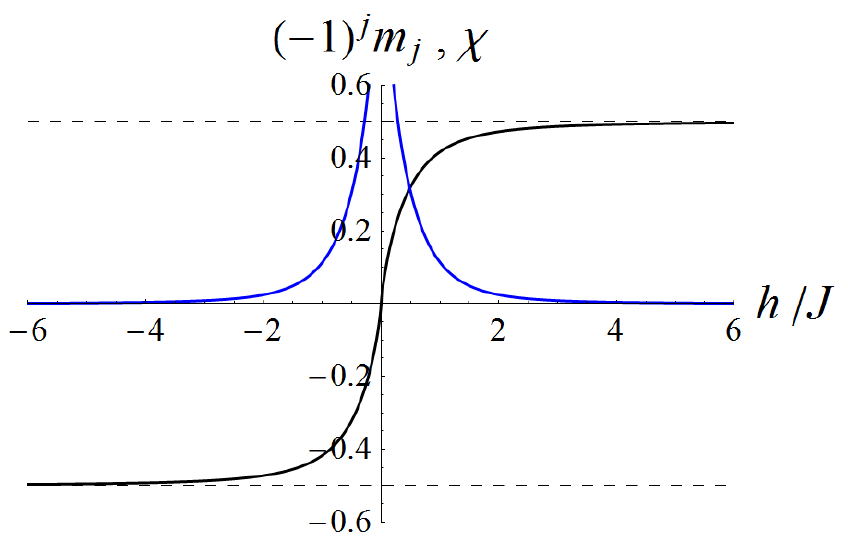}
	\caption[Staggered magnetization and the susceptibility]{\footnotesize (Color online) Staggered magnetization $(-1)^j m_j$, passing through 0 at $h=0$ and saturating at $\pm \frac{1}{2}$ for $|h|\rightarrow\infty$, 
 and susceptibility $\chi$, diverging for $h\rightarrow 0$.}
	\label{fig:staggered}
\end{figure}

For a staggered field, $h_j=(-1)^j h, j=1,...,L$, the single-particle spectrum has a gap $2\vert h\vert$ separating a valence from a conduction band for an arbitrarily small field amplitude $h$. Moreover there is long-range antiferromagnetic order with local moments given by
\be
m_j = \frac{(-1)^j}{\pi}\frac{1}{\sqrt{1+\lambda^{2}}}\,\mathcal{K}\left(\frac{1}{1+\lambda^2}\right)\, ,
\label{magnetization_staggered}
\ee
where $\mathcal{K}(x)$ is a complete elliptic integral of the first kind and $\lambda=h/J$.
Fig. \ref{fig:staggered} shows both the order parameter (the staggered magnetization 
$(-1)^jm_j$) and the susceptibility $\chi=(-1)^j\, dm_j/dh$. The order parameter grows first linearly as a function of the field and saturates for $|h|\gg J$. The susceptibility diverges at the critical point $h=0$.

The ground state can also be readily obtained if both homogeneous and staggered fields are included. Besides phases with fully aligned or alternating spins a third phase is found where both orders coexist \cite{Alcaraz}.

From now on we will mainly concentrate on interfaces, such as the step model or odd-numbered rings with staggered transverse fields. Sometimes more general inhomogeneous field patterns will be considered, subject to the constraint 
\begin{equation}
\sum_{j=1}^L h_j=0.
\label{constraint}
\end{equation} 
This condition of vanishing average field will also be applied to the two particular configurations of Fig. \ref{fig:fields}, where it holds 
automatically for even-numbered chains. To satisfy the constraint for an odd number of sites
we assume the field to vanish at a single site (at the site of the interface for the step model), 

Instead of inhomogeneous fields one could also consider inhomogeneous exchange constants $J_j$. Their effects are however expected to be much weaker than those of an inhomogeneous field. Consider an even-numbered chain with exchange constants varying arbitrarily in sign, but not in absolute value. Applying the canonical transformation $S_j^x\rightarrow-S_j^x$, 
$S_j^y\rightarrow-S_j^y$ and $S_j^z\rightarrow S_j^z$ at appropriate places then leads to a model with homogeneous exchange. In contrast, one cannot simply gauge away varying signs of transverse fields without modifying other terms in the Hamiltonian. 

The paper is organized as follows. Symmetry considerations are presented in Section II, which are very useful for the XX chain with interface, despite the fact that translational invariance is explicitly broken. For general fields vanishing on average the concept of particle number parity can be used to determine the ground state degeneracies. For an even number of sites and more restricted field patterns, namely those that are odd under reflection about the chain center,  we find a particular 
SU(2) symmetry involving operators $O_x, O_y, O_z$. While $O_z$ is essentially equal to the 
$z$ component of the total spin, the other two operators generate mappings consisting of both a reflection and an electron-hole transformation. Section III discusses the step model for an even-numbered chain. The critical point separating a ``metallic'' from an ``insulating'' phase is at $|h|=J$, in close analogy to the Lifshitz transition for a homogeneous field. For $|h|<J$, pronounced Friedel oscillations in the local moments (or the particle density) occur. Both their characteristic wavelength and their intensity depend on the field strength. Chapter IV deals with
the intriguing case of odd-numbered chains, both for staggered fields and for the step model. In both cases an isolated level appears in the gap, which is not exactly at zero energy. Nevertheless the ground state energy does not depend on the occupancy of this level, in perfect agreement with the symmetry arguments of Section II. Although the SU(2) symmetry of even chains no longer holds, the unitary transformations introduced in Section II can be generalized to odd chains. One of these transformations links the two degenerate ground states and has the form of a Majorana operator. A coherent superposition of these two states, a possible qubit, has a magnetic moment in the plane, with some angle and an amplitude $\le\frac{1}{2}$. Section V presents a brief summary as well as a discussion of possible extensions, such as the XXZ model or the XY chain with anisotropic exchange.

\section{Symmetry considerations}
\subsection{Particle-number parity}
The Hamiltonian (\ref{hamiltonian_2}) depends on the particle-number parity $P=\exp (i\pi N)$, a concept that has been successfully applied  in other contexts, such as  even-odd effects in small
metallic grains \cite{Halperin}. This operator commutes with the Hamiltonian and therefore both the single-particle levels and the many-particle eigenstates can be grouped in two parity sectors according to the eigenvalues of $P$ ($\pm 1$). For transverse fields satisfying the constraint (\ref{constraint})
the Hamiltonian has also particle-hole symmetry, in the following sense. The transformation $c_j\rightarrow c_j^\dag$, $c_j^\dag\rightarrow c_j$, $j=1,...,L$, simply changes the sign of the first two terms of the Hamiltonian. As to the boundary term, one has to take into account that at the same time $N$ is transformed to $L-N$ and therefore $P\rightarrow P$ for even $L$ and $P\rightarrow -P$ for odd $L$. Therefore this particle-hole transformation amounts to the mapping
\be
(H,P)\rightarrow\left\{\begin{array}{ll}(-H,P),&L\quad \mbox{even},\\(-H,-P),&L\quad \mbox{odd}.\end{array}\right.
\ee
Because this is a canonical transformation, the spectrum does not change. For even
chains the energy levels occur in pairs $(\varepsilon_\nu, -\varepsilon_\nu)$ in both parity sectors. For odd chains the energy level $\varepsilon_\nu$ of one parity sector has its partner 
$-\varepsilon_\nu$ in the opposite parity sector. To determine the ground state one has to explore both sectors. Setting aside the ambiguity of zero-energy levels, one quickly realizes that the rule of simply occupying all the negative-energy levels is not always consistent. In fact, the calculation of the single-particle spectrum for a given parity sector may produce a number of negative-energy levels that disagrees with the parity chosen initially. For even chains the number of negative-energy levels is equal to $L/2$, in both parity sectors, and one expects the ground state to have parity $P=\exp{(i\pi L/2)}$, for which it is consistent to occupy the negative-energy levels and to leave empty the positive-energy levels. Thus the ground state is expected to be unique for even chains. 

For odd chains the situation is slightly more complicated. If there are $N$ negative-energy levels in one sector, there are $L-N$ negative energy levels in the other, but this number $N$ may or may not agree with parity, as will become more clear for the explicit examples of Section IV. Let us first discuss the case where parity and number of negative-energy levels are consistent. The ground state energy in the + parity sector is then given by
\be
E^{(+)}=\sum_{\nu, \varepsilon_\nu^{(+)}<0}\varepsilon_\nu^{(+)}\, .
\ee
Together with the relation $\sum_\nu\varepsilon_\nu^{(+)}=0$, a simple consequence of the constraint $\sum_j h_j=0$, and the fact that to each energy level $\varepsilon_\nu^{(+)}$ there exists a level $\varepsilon_\nu^{(-)}=-\varepsilon_\nu^{(+)}$, we obtain
\be
E^{(+)}=-\sum_{\nu, \varepsilon_\nu^{(+)}>0}\varepsilon_\nu^{(+)}
=\sum_{\nu, \varepsilon_\nu^{(-)}<0}\varepsilon_\nu^{(-)}=E^{(-)}\, .
\ee
Therefore there are two orthogonal states with the same energy. If the particle-number parity and the number of negative-energy levels are not consistent, one has either to add a particle to the lowest unoccupied positive-energy level or to remove a particle from the highest occupied negative-energy level. Particle-hole symmetry together with the vanishing trace of eigenvalues can then again be used to show that the many-particle states in the two sectors have the same energy. We conclude that the ground state for an odd-numbered chain is degenerate. 

\subsection{SU(2) symmetry ($L$ even)}
Consider an even-numbered chain, $L=2M$, with fields that are odd under reflection about the chain center, i.e.
\be
h_j=-h_{L+1-j},\quad j=1,...,M.
\label{fields_reflection}
\ee
The two field patterns of Fig. \ref{fig:fields} satisfy this relation. We now introduce the operators
\begin{eqnarray}
O_x&=& \frac{1}{\sqrt{2}}\sum_{j=1}^M(-1)^j(c_{L+1-j}c_j+c_j^\dag c_{L+1-j}^\dag)\, ,\nonumber\\
O_y&=& \frac{i}{\sqrt{2}}\sum_{j=1}^M(-1)^j(c_{L+1-j}c_j-c_j^\dag c_{L+1-j}^\dag)\, ,\nonumber\\
O_z&=&\sum_{j=1}^L(c_j^\dag c_j-\frac{1}{2})\, .
\label{pairingop}
\end{eqnarray}
They satisfy the commutation relations of an angular momentum
\begin{equation}
[O_x,O_y]=iO_z,\, [O_y,O_z]=iO_x,\, [O_z, O_x]=iO_y
\end{equation} 
and commute with the Hamiltonian (\ref{hamiltonian_2}) for field configurations satisfying 
Eq. (\ref{fields_reflection}). Therefore our model has SU(2) symmetry. We note that the XY  Hamiltonian (\ref{hamiltonian_2}) has of course not the SU(2) symmetry with respect to the true spin operators. Rather the operators (\ref{pairingop}) are similar to the $\eta$ spin introduced by Yang and Zhang in the case of the Hubbard model \cite{Yang_90}. The possible eigenvalues of
$O_z$ are integers between $-M$ and $+M$. Thus the many-particle states may have rather large degeneracies,
in contrast to the single-particle states, which are expected to be non-degenerate.
Certain (many-particle) eigenstates can then be constructed in a similar way as those of the ordinary angular momentum operator. For instance, starting with the eigenstate for $N=L$, where all sites are occupied, we can generate an eigenstate for $N=L-2$ by applying the lowering operator
\begin{equation}
O_-=O_x-iO_y=\sqrt{2}\sum_{j=1}^M(-1)^jc_{L+1-j}c_j\, .
\end{equation}
In a similar way one finds an eigenstate for $N=2$ by applying
\begin{equation}
O_+=O_x+iO_y=\sqrt{2}\sum_{j=1}^M(-1)^jc_j^\dag c_{L+1-j}^\dag
\end{equation}
to the ground state for $N=0$, where all sites are unoccupied.
\subsection{Symmetry transformation}
The operators introduced above can be used as generators of symmetry transformations. As a particular example we consider the unitary operator
\begin{equation}
U=e^{-i\frac{\pi}{\sqrt{2}}O_x}\, ,
\label{unit_tr}
\end{equation}
which induces a particle-hole transformation plus reflection, 
\begin{eqnarray}
Uc_jU^\dag &=& i(-1)^jc^\dag_{L+1-j}\, ,\nonumber\\
Uc_{L+1-j}U^\dag &=& i(-1)^jc^\dag_j\, ,
\label{ph_transform}
\end{eqnarray}
where $j=1,...,M$ and $L=2M$. It leaves the Hamiltonian (\ref{hamiltonian_2}) invariant if 
Eq. (\ref{fields_reflection}) holds. One also finds the simple relation
\be
U^2=\prod_{j=1}^M[2(n_j-n_{L+1-j})^2-1]\, ,
\label{usquared}
\ee
where $n_j=c_j^\dag c_j$. This operator is both hermitian and unitary and therefore its possible eigenvalues are $\pm 1$. As an example we choose a state with an even number of particles distributed over the $L$ sites such that $n_j=n_{L+1-j}, \, j=1,...,M$. In this case $U^2=(-1)^M$. It is easy to check that any permutation of particles does not change this value and therefore any state of an even number of particles (or of even particle number parity) is an eigenstate of $U^2$ with eigenvalue $(-1)^M$. Similarly, for an odd number of particles we can start with a configuration where the relation $n_j=n_{L+1-j}, \, j=1,...,M$ holds for all sites except one. Therefore one of the factors in Eq. (\ref{usquared}) is equal to $1$ whereas all the others are -1. The eigenvalue of $U^2$ for any state with odd number parity is $-(-1)^M$. Thus any $N$-particle state of a chain with $L=2M$ is an eigenstate of $U^2$ with eigenvalue $(-1)^{N+M}$. We conclude that up to a phase factor of $(-1)^M$ the operator $U^2$ is identical to the particle number parity $P$.
\section{Friedel oscillations ($L=2M$)}
The step model of Fig. \ref{fig:fields} can be viewed as the union of two chains with different homogeneous transverse fields. If interface effects are neglected, each part will undergo separately a quantum phase transition, one at the left critical point of Fig. \ref{fig:homogeneous}, the other at the right point. This can be quantified as follows. We introduce fermionic operators $a_j, b_j$ through
\be
c_j=\left\{\begin{array}{ll}a_j,&j=1...,M\\ b_j,&j=M+1,...,L\end{array}\right.
\ee
and write
\beq
H&=&\sum_{j=1}^M\left\{\frac{J}{2}(a_j^\dag a_{j+1}+b_j^\dag b_{j+1}
+\mbox{ h.c.})\right.\nonumber\\&&\qquad\left. +h(a_j^\dag a_j -b_j^\dag b_j)\right\}+H'\, ,
\label{hamiltonian_3}
\eeq
where $H'$ consists of boundary terms proportional to $J$ and $a_{M+1}=a_1$, $b_{M+1}=b_1$. If we neglect $H'$ we find two independent chains with homogeneous fields $-h$ and $+h$, respectively, and with single-particle spectra $J\cos{k}\pm h$. For $|h|>J$ one band is empty, the other one is full. In this limit, the magnetic moments $m_j$ are $-\frac{1}{2}$ for $j=1,...,M$ and $+\frac{1}{2}$ for $j=M+1,...,L$.

\begin{figure}[ht]
	\centering
	\includegraphics[width=8cm]{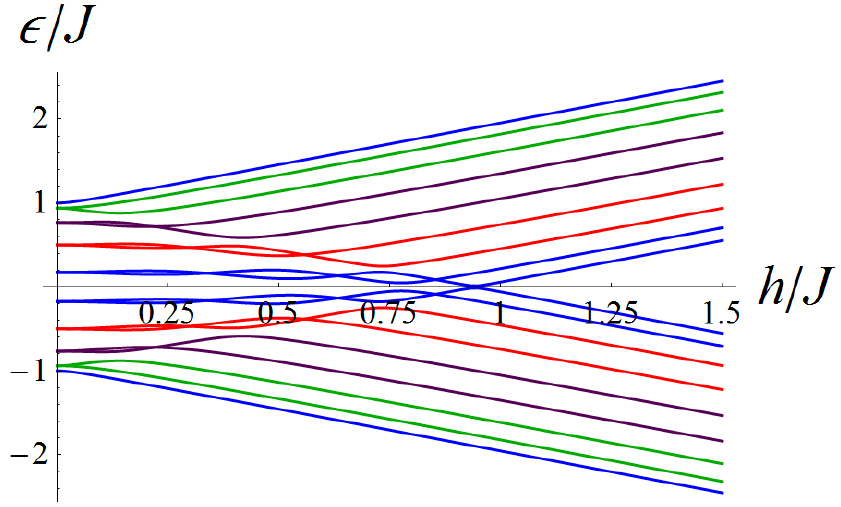}
	\caption[Single-particle spectrum of the interface case]{\footnotesize  (Color online) Energy levels of the step model for $L=18$. An energy gap opens for 
$h/J>1$, signaling the metal-insulator transition.}
	\label{fig:spectrum_step}
\end{figure}

\begin{figure}[ht]
	\centering
	\includegraphics[width=8cm]{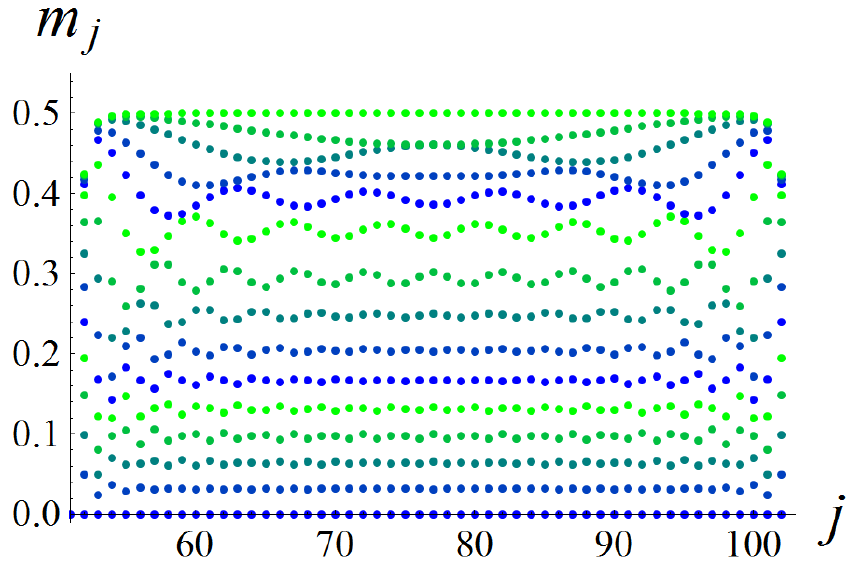}
	\caption[Local magnetization of the interface case]{\footnotesize  (Color online) Magnetic moments $m_j$ of the interface model for $L=102$ and $j=51,...,102$. The different lines correspond to different field strengths; these are, from bottom to top, $h=0.00, 0.10,..., 0.90, 0.95, 0.97, 0.99,0.993, 0.999$.}
	\label{fig:magnetization_Leven_simple_selection}
\end{figure}

The boundary terms in $H'$ are negligible for $\vert h\vert\gg J$, but not for 
$\vert h\vert<J$. To study the ``metallic'' region we have numerically diagonalized the Hamiltonian.
The single-particle spectrum is illustrated in Fig. \ref{fig:spectrum_step}. A gap exists for $|h|>J$, while the system is gapless for $|h|<J$, as expected.  The local moment at site $j$ is given by
\be
m_j=-\frac{1}{2}+\sum_{\nu, \varepsilon<0} |u_{j\nu}|^2\, ,
\ee
where $u_{j\nu}$ are the matrix elements of the unitary transformation that diagonalizes the Hamiltonian ($\nu$ numbers the energy levels).
The numerical results are illustrated in Fig. \ref{fig:magnetization_Leven_simple_selection} for several values of the field strength. Only the right half of the chain is represented, the left segment being simply obtained by inversion with respect to the origin of the axes. Pronounced oscillations are observed with amplitudes that first increase as $h$ increases and then quickly disappear at the approach of the critical field strength, where the magnetization saturates at $m_j=+\frac{1}{2}$ in the right half of the chain (and $-\frac{1}{2}$ to the left). 
Both for $h\ll J$ and for $h\gg J$ the magnetic moments can be calculated using perturbation theory, as shown in Appendix \ref{perturbation_theory}. In the appropriate limits the results agree well with those shown in Fig. \ref{fig:magnetization_Leven_simple_selection}. Thus for $h\gg J$ the expansion in powers of $J/h$ shows that the mutual influence of the two
different orderings is only effective in close proximity to the interface.

To make contact with the conventional picture of Friedel oscillations, we split the Hamiltonian into two parts as in Eq. (\ref{hamiltonian_3}) and consider field strengths smaller than $J$. The first term represents two metallic systems with Fermi wave vectors $\pm \arccos{(h/J)}$. The surface term $H'$ is then expected to produce Friedel oscillations with period $\pi/\arccos{(h/J)}$. This prediction is in perfect agreement with the periods of the oscillations seen in the figure. A closer look shows that for some specific field strengths the oscillations are practically absent, while they are particularly pronounced for others. We attribute this effect to interference between Friedel oscillations originating from the two interfaces located, respectively, in the middle and at the end of the chain.

\section{Odd-numbered chains}

We turn now to the discussion of odd-numbered chains. To preserve the condition $\sum_jh_j=0$, we start from an even-numbered chain and add a site at which the magnetic field vanishes. The choice of the site is arbitrary for the staggered field while for the step model we insert it at the interface. Choosing in both cases the additional site at $\ell=(L+1)/2$, we can write the field term of the Hamiltonian as 
\be
-h\sum_{j=1}^{\ell-1}\sigma_j(c_j^\dag c_j-c_{L+1-j}^\dag c_{L+1-j})\, ,
\ee
where $\sigma_j=(-1)^j$ for a staggered field and $\sigma_j=1$ for the step model. 
\subsection{Midgap level}
To understand qualitatively the spectrum, we consider first the limit $|h|\gg J$ and bypass the new site by replacing the hopping terms involving the site $\ell$ by a single hopping between $\ell -1$ and $\ell +1$. We obtain two independent systems, an even chain -- with either a staggered field or a field step -- plus a single site without field. In this limit the spectrum consists of a filled valence band, an empty conduction band and a single midgap level at $\varepsilon=0$. 

\begin{figure}[ht]
	\centering
	\includegraphics[width=8cm]{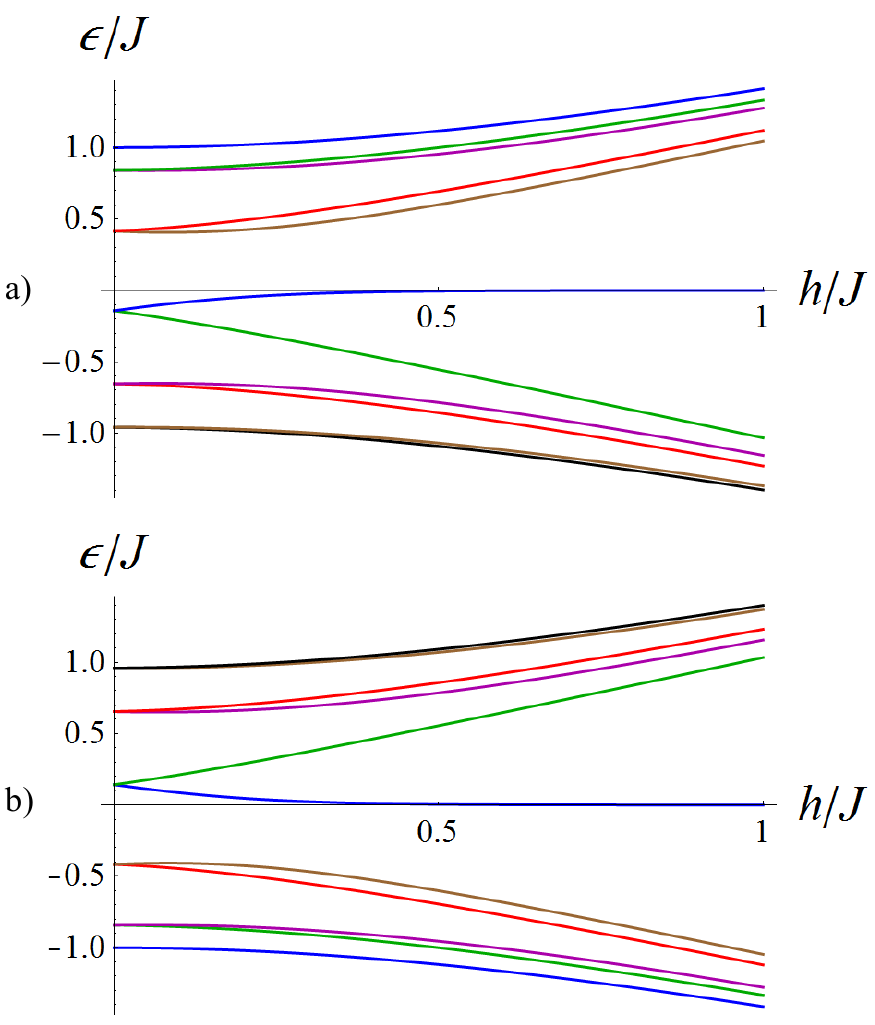}
	\caption[Single-particle spectrum with both parity, L=11 + Staggered field]{\footnotesize  (Color online) Single-particle spectrum of the odd-numbered chain $L=11$ with a staggered field. Both parity sectors are shown, in a) $N$ is odd while in b) $N$ is even. Note that all energy levels are non-degenerate; for one of the outermost levels the splitting is barely visible.}
	\label{fig:spectrum_Lodd_staggered_L11_bothN}
\end{figure}

The numerical diagonalization of the Hamiltonian confirms this picture. Fig. \ref{fig:spectrum_Lodd_staggered_L11_bothN} shows the result for a chain with 11 sites and a staggered field at all but one site. The spectrum clearly exhibits the two bands plus an isolated level, which tends very quickly to midgap as $h$ increases. The two figures correspond to two different particle numbers. $N$ is odd in the upper case and even in the lower case. This is at odds with the number of negative-energy levels, which is even in the upper case and odd in the lower case. Therefore in the ground state the midgap level is not occupied in the upper case, although its energy is slightly negative, while it is occupied in the lower case, although its energy is positive. The particle-hole symmetry between the two figures is obvious. It implies, as explained in Section II, that the total energies of the two distinct states are equal, hence the ground state is doubly degenerate. For the step model similar results are obtained for $\vert h\vert >J$.

In the presence of an energy gap of size $2\Delta$ the wave function $u_{j0}$ of the midgap state is localized, i.e. it decreases as $\exp{(\vert j-\ell\vert /\xi)}$ with a localization length 
$\xi\approx J/\Delta$. For the step model (in the limit $L\rightarrow\infty$) $\xi$ diverges at the critical point. 

\subsection{Symmetry transformation}
The operators (\ref{pairingop}) can also be used for odd-numbered chains, with $M=\ell-1$. They no longer commute with the Hamiltonian, but the modified unitary transformation
\be
U'=e^{-i\frac{\pi}{\sqrt{2}}O_x}(c_\ell +c_\ell^\dag)
\label{unit_tr2}
\ee
together with a local gauge transformation does. To see this, we first notice that the relations 
(\ref{ph_transform}) remain valid for $j=1,...,\ell-1$. Moreover we find
\be
U'c_\ell U'^\dag = c_\ell^\dag\, .
\ee
The application of this transformation together with the replacement
$c_\ell\rightarrow i(-1)^\ell c_\ell$ indeed leaves the Hamiltonian invariant. 

In contrast to the operator $U$ of Eq. (\ref{unit_tr}), which is bosonic, the operator $U'$ is fermionic. 
Moreover it satisfies the relations
\be
U'^\dag=U'(-1)^{M+N} ,\quad U'^2=(-1)^{M+N}
\ee
and thus is similar to the creation (or annihilation) operator of a Majorana fermion.
\subsection{Degenerate ground state}
We show now that the members of the ground state doublet are transformed into each other by the unitary operator $U'$.
We use the notation $|\pm\rangle$ for the two states which are eigenstates of $S^z$ with eigenvalues $\pm 1/2$. In fact, in view of Eq. (\ref{spin}) we get
\be
S^ z|\pm\rangle=(N-\frac{L}{2})|\pm\rangle=\pm\frac{1}{2}|\pm\rangle\, .
\label{eigenstate}
\ee 
Because $U'$ commutes with the Hamiltonian, the state $U'^\dag|+\rangle$ has the same energy eigenvalue as the state 
$|+\rangle$ and must be a superposition of $|+\rangle$ and $|-\rangle$. Moreover the relation $U'S^zU'^\dag=-S^z$ implies 
\be
S^zU'^\dag|+\rangle = -U'^\dag S^z|+\rangle=-\frac{1}{2}U'^\dag |+\rangle
\ee
and therefore we can identify $U'^\dag|+\rangle$ with $|-\rangle$. We conclude that the two degenerate states $|\pm\rangle$ are transformed into each other as
\be
|-\rangle=U'^\dag|+\rangle\, ,\qquad |+\rangle= U'|-\rangle\, .
\ee

An arbitrary superposition of these two states is also an eigenstate of the Hamiltonian. We may define a qubit as the state
\be
|\beta\rangle=\frac{1}{\sqrt{2}}(|+\rangle+e^{i\beta}|-\rangle)\, .
\label{qubit}
\ee
The expectation value of $S^z$ for this state vanishes, but not that of $S^ x$ or $S^ y$. To estimate 
$\langle\beta|S^\alpha|\beta\rangle$, $\alpha=x, y$, we use the limit
$|h|\gg J$, for which we can cut the connections between the added site and the rest of the chain (the neglected terms are of order $J$). We obtain the factorization
\be
|+\rangle=|\Phi\rangle\otimes|1\rangle\, , \qquad |-\rangle=|\Phi\rangle\otimes|0\rangle\, ,
\ee
where $|1\rangle$ and $|0\rangle$ are the two possible states of the added site with occupation numbers 1 and 0, respectively, and $|\Phi\rangle$ is the ground state for the Hamiltonian involving all the other sites. In this approximation $|\Phi\rangle$ does not depend on the occupation of the added site. It is convenient to split the spin operators as follows
\be
S^\alpha=S'^\alpha+S_\ell^\alpha\, ,
\ee
where $\ell$ is the added site and
\be
S'^\alpha=\sum_{j\neq\ell}S_j^\alpha\, .
\ee
Because of Eq. (\ref{eigenstate}) $|\Phi\rangle$ can be assumed to be an eigenstate of $S'^z$  with eigenvalue zero. It follows that also the expectation values of $S'^x$ and $S'^y$ with respect to $|\Phi\rangle$ vanish. It is then easy to see that for the qubit state $|\beta\rangle$ we obtain the simple result
\be
\langle\beta| S^x|\beta\rangle= \frac{1}{2}\cos{\beta}\, ,\qquad 
\langle\beta| S^y|\beta\rangle= \frac{1}{2}\sin{\beta}\, .
\ee

In the limit $|h|\gg J$ the double degeneracy originates from the existence of a zero-energy level, which may or may not be occupied. For smaller fields
the degeneracy is not simply related to different possible occupancies of the zero-energy level, but relies on global symmetries. This is quite different from the famous midgap states associated with kink solitons in the SSH model \cite{Heeger}, where the zero-energy level can be empty, singly or doubly occupied, and where neither the midgap state nor the band states are affected by a change in occupancy. In our case a change in the particle number by 1 modifies the Hamiltonian, and therefore both the midgap state and the band states are altered. 

It is remarkable that the degeneracy of the ground state does not depend on the field configuration except that the trace is required to vanish, $\sum_jh_j=0$. If this condition is fulfilled, field fluctuations will not lead to decoherence. Moreover, for the two examples studied here the doubly degenerate ground state is separated from the excited states by a finite gap, provided that the field strength is large enough in the case of the step model ($|h|>J$). Thus the conditions for a useful qubit, low decoherence and excitation gap, seem to be well satisfied by our model.  Both the relation (\ref{eigenstate}) and the subsequent discussion remain valid, except that now there is no
guarantee for the existence of an energy gap. Therefore the state (\ref{qubit}) cannot be used
indiscriminately as a well protected qubit.
\subsection{Generalization}
We can easily extend these considerations to more general field patterns, namely to those defined by Eq. (\ref{fields_reflection}) for $j=1,...,\ell-1$ together with the condition $h_\ell=0$. For simplicity we use open boundary conditions, for which the parity-dependent term in the Hamiltonian 
(\ref{hamiltonian_2}) is absent. In this case the transformation 
$c_j \leftrightarrow (-1)^jc_{L+1-j},\, j=1,...,\ell,$ maps $H$ into $-H$, and the single-particle spectrum is symmetric, i.e. energy levels occur in pairs $(\varepsilon_\alpha, -\varepsilon_\alpha)$, except for one level which is mapped onto itself and therefore has zero energy. In one of the ground states the zero-energy level is empty, in the other it is occupied; they correspond to the states 
$\vert +\rangle$ and $\vert -\rangle$ described above. 
\section{Discussion}
In this paper we have studied the ground state of the XX chain with an interface defined by specific patterns of the transverse field. Two types have been investigated in detail, a staggered field where an interface is produced automatically for odd-numbered rings and a step model where the field jumps suddenly from a negative value $-h$ to a positive value $+h$. 
The step model illustrates nicely the effects of an interface separating two halves with different magnetizations, or with different particle densities in the fermionic language. In the ``metallic'' regime, $\vert h\vert <J$, we have found Friedel oscillations which become quite pronounced with increasing field strength but then quickly die out as the critical point $\vert h\vert =J$ is approached. For larger fields a proximity effect remains, limited to the sites adjacent to the interface. To realize the step model, or at least a smooth step satisfying the condition (\ref{fields_reflection}), one could imagine two identical U-shaped magnets placed symmetrically with respect to the center, but with north and south poles interchanged. 

Odd-numbered rings exhibit some rather intriguing phenomena. For any field pattern satisfying the relation $\sum_j h_j=0$ the ground state is doubly degenerate. In the insulating phase of our two interface models this degeneracy is linked to a level close to the center of the energy gap.
For finite lengths the precise location of this level shifts as the particle-number parity $P=\exp{(i\pi N)}$ is changed, due to a parity-dependent boundary term. Thus the degeneracy for an odd number of sites originates from a global symmetry and not simply from the occupancy of a zero-energy level. An explicit unitary transformation $U'$ has been constructed, which maps one ground state to the other. The operator $U'$ has the properties of a creation (or annihilation) operator of a Majorana fermion, and $U'^2$ is essentially equal to the parity $P$. The doubly degenerate ground state has spin $S_z=\pm \frac{1}{2}$ and may be a promising candidate for a well protected qubit.

An interesting question is to what extent the present results survive if the coupling between $z$ components of the spins is also included. In the fermionic representation it amounts to considering an interaction term between particles on nearest-neighbor sites.  
This problem has been studied numerically both for homogeneous and staggered fields \cite{Alcaraz}, but to our knowledge not in the presence of an interface. The coupling between $z$ components breaks the SU(2) symmetry established in Section II, yet it does not spoil the invariance of the Hamiltonian under the transformations (\ref{unit_tr}) and (\ref{unit_tr2}) for even and odd chains, respectively. Therefore the eigenstates of odd-numbered chains remain degenerate.

A further possible generalization would be anisotropic exchange, $J_x\neq J_y$. In this case the Jordan-Wigner transformation leads to a BCS-type model and, as shown by Kitaev \cite{Kitaev}, to the possibility of Majorana fermions at the two ends of a chain in a (homogeneous) transverse field.

\acknowledgments
We have profitted from stimulating discussions with George Japaridze, Eduardo Marino and Pasquale Sodano. One of us (G.F.) thanks the Fribourg Center for Nanomaterials for support during his Master studies.
\appendix
\section{Solution of the XX chain for homogeneous and staggered transverse fields}
\label{homogeneous_staggered}
\subsection{Homogeneous field}
 For odd $N$ and a constant transverse field, $h_j=h, j=1,...,L$, the Hamiltonian 
 (\ref{hamiltonian_2}) is diagonalized by Fourier transformation 
\be
c_j=\frac{1}{\sqrt{L}}\sum_ke^{ikj}c_k\, ,
\label{Fourier}
\ee
where $k=\frac{2\pi}{L}\nu$, $-L/2<\nu\le L/2$. We find 
\be
H=\sum_k(J\cos{k}-h)c_k^\dag c_k\, .
\ee
For even $N$, the undesirable sign of the boundary terms can be distributed in a homogeneous way over the entire ring using the transformation
$c_j\rightarrow e^{i\varphi j}c_j$. where $\varphi = \pi/L$. The transformed Hamiltonian
\be
H=\frac{J}{2}\sum_{j=1}^L(e^{i\varphi}c_j^\dag c_{j+1}+\mbox{h.c.})
-h\sum_{j=1}^L c_j^\dag c_j
\ee
can then again be diagonalized by Eq. (\ref{Fourier}), leading to
\beq
H&=&\sum_k[J\cos{(k+\varphi)}-h]\, c_k^\dag c_k\nonumber\\
&=&{\sum_k} '(J\cos{k}-h)\, c_k^\dag c_k\, ,
\eeq
where in $\sum'_k$ the $k$ values are shifted with respect to those in $\sum_k$; they assume values 
$\frac{2\pi}{L}(\nu+\frac{1}{2})$, $-L/2\le\nu< L/2$.
\subsection{Staggered field ($L$ even)}
For a staggered transverse field, $h_j=(-1)^jh$, $j=1,...,L$, and an even number of sites the Hamiltonian has full particle-hole symmetry. For simplicity we choose a chain length $L=4n+2$, for which the ground state is non-degenerate with an odd number of particles, $N=2n+1$. The Hamiltonian then reads

\beq
H&=& \frac{J}{2}\sum_{j}(c^{\dagger}_{j}c_{j+1} + \mbox{h.c.}) 
- h\sum_{j}{(-1)}^{j}c^{\dagger}_{j}c_{j}\nonumber \\
&=&\sum_{k}(\varepsilon_k c_k^\dag c_k
-h\, c_k^\dag c_{k\pm\pi})\, ,
\label{hamiltonian_staggered}
\eeq
where $\varepsilon_k=J\cos{k}$. 
The Bogoliubov transformation ($k>0$)
\begin{eqnarray}
&c_{k}& = ~~\cos{\vartheta_{k}}\,\gamma_{k} + \sin{\vartheta_{k}}\,\gamma_{k-\pi}\, , \\
&c_{k - \pi} & = -\sin{\vartheta_{k}}\,\gamma_{k} + \cos{\vartheta_{k}}\,\gamma_{k-\pi}~,
\end{eqnarray}
diagonalizes the Hamiltonian (\ref{hamiltonian_staggered}), provided that $\tan{2\vartheta_k}=h/\varepsilon_k$.
Thus we get
\be
H=\sum_kE_k\gamma_k^\dag\gamma_k\, ,
\ee
where the single-particle spectrum is given by
\be
E_k=\mbox{sign}(\varepsilon_k)\sqrt{\varepsilon_k^2+h^2}\, .
\ee
The local moment $m_j=\langle S_j^z\rangle$ is found to be
\be
m_j=  \frac{{(-1)}^{j}h}{L}\sum_{k>0}\frac{1}{|E_k|}\,.
\ee
For $L\rightarrow\infty$ we replace the $k$ sum by an integral and obtain
Eq. (\ref{magnetization_staggered}).
\section{Perturbation theory for the step model}
\label{perturbation_theory}
We consider the step model, $h_j=-h$ to the left and $h_j=h$ to the right.
For simplicity we choose $L=2M$ with $M$ odd. The ground state then has $N=M$ particles and the hopping term in Eq. (\ref{hamiltonian_2}) has the conventional tight-binding form.
In two limiting cases, $|h|\ll J$ and $|h|\gg J$, we can calculate interface effects perturbatively. In these two limits, the Hamiltonian (\ref{hamiltonian_2})
is split in different ways into a dominant part $H_0$ and a perturbation $H'$.
For $|h|\ll J$ we make the obvious choice
\beq
H_0&=&\frac{J}{2}\sum_{j=1}^L(c_j^\dag c_{j+1}+\mbox{h.c.})
=\sum_k\varepsilon_k\ c_k^\dag c_k\, ,\nonumber \\
H'&=&h\sum_{j=1}^M( c_j^\dag c_j-c_{L+1-j}^\dag c_{L+1-j})\nonumber\\
&=&\sum_{k,k'}\tilde{h}_{k-k'}\ c_k^\dag c_{k'}\, ,
\eeq
where
\beq
\varepsilon_k&=&J\cos{k}\, ,\nonumber\\
\tilde{h}_q&=&-ih(1-\cos qM)\frac{e^{-iq/2}}{\sin(q/2)}\, ,
\eeq
while for $|h|\gg J$ we just interchange the two terms. Therefore the unperturbed ground states $|\Phi_0\rangle$ are, respectively, the half-filled band 
(levels with $|k|<\frac{\pi}{2}$ occupied, the others not) and 
the half-filled chain (left sites occupied, right sites empty). 
Using the fundamental relation of many-body perturbation theory, 
we can obtain the local moment
$m_j=\langle c_j^\dag c_j\rangle -\frac{1}{2}$ from
\be
\langle c_j^\dag c_j\rangle=\langle\Phi_0|T\{ c_j^\dag c_j\ U(\infty,-\infty)\}
|\Phi_0\rangle_{\mbox{\small con}}\, ,
\label{many-body}
\ee
where 
\be
U(\infty,-\infty)=\exp\left\{{-\frac{i}{\hbar}
\int_{-\infty}^{+\infty}dt\ H'(t)}\right\}\, ,
\ee
$T$ is the time-ordering operator and only connected diagrams contribute.

We discuss first the case of weak fields, $|h|\ll J$. In the unperturbed 
ground state the magnetic moments $m_j$ vanish. The first-order 
correction is found to be
\be
m_j^{(1)}=\frac{1}{L^2}\sum_{k,k'}\ e^{i(k-k')j}\ \tilde{h}_{k-k'}
\frac{n_k-n_{k'}}{\varepsilon_k-\varepsilon_{k'}}\, ,
\ee
where $n_k$ is the Fermi distribution of the unperturbed ground state. In the thermodynamic limit the sums are replaced by integrals. Introducing the variables $Q=\frac{1}{2}(k+k')$ and $q=k-k'$, we can perform one integration and obtain
for the region close to the interface located between sites $M$ and $M+1$
\be
m_{M+\ell}=\frac{h}{\pi^2J}(-1)^\ell\int_0^\pi dq\ 
\frac{\cos{q(\ell-\frac{1}{2})}}{\cos^2{\frac{q}{2}}}\ 
\log(\tan\frac{q}{4})\, .
\label{asymptotic}
\ee
The remaining integral is readily computed numerically. One obtains a step-like contribution corresponding to average positive and negative magnetizations to the left and right of the interface, respectively, plus an alternating part 
(wave vector $\pi$) which decays like $1/|\ell|$ away from the interface. For sites close to the other interface (located between the connected chain ends) the pattern (\ref{asymptotic}) is simply multiplied by a minus sign. This perturbative formula agrees well with numerical results for $h<0.2J$. For larger values of $h$ the dominant wave vector is no longer $\pi$, but $\pi/\arccos{(h/J)}$, as explained in Section IV.

We now discuss the other limit, $h\gg J$, where the hopping term is the perturbation. In the unperturbed ground state the sites to the left of the interface are occupied while those to the right are empty, corresponding to magnetic moments $+\frac{1}{2}$ and $-\frac{1}{2}$, respectively. It is easy to see that the first order correction vanishes. Up to second order we find
\be
m_{M+\ell}=\left\{\begin{array}{ll}
\frac{1}{2},&\ell\le -1,\\
\frac{1}{2}-\left(\frac{J}{4h}\right)^2,&\ell=0,\\
-\frac{1}{2}+\left(\frac{J}{4h}\right)^2,&\ell=1,\\
-\frac{1}{2},&\ell\ge 2.
\end{array}\right.
\ee
\vskip 1cm

\

. 
\end{document}